\documentclass[reprint,nofootinbib,superscriptaddress, aps,longbibliography]{revtex4-1}
\usepackage{amsmath}
\usepackage{braket}
\usepackage{graphicx}
\usepackage{color}
\usepackage{footnote}
\usepackage[table]{xcolor}

\newwrite\bibnotes
    \def\bibnotesext{Notes.bib}
    \immediate\openout\bibnotes=\jobname\bibnotesext
    \immediate\write\bibnotes{@CONTROL{REVTEX41Control}}
    \immediate\write\bibnotes{@CONTROL{%
    apsrev41Control,author="08",editor="1",pages="1",title="0",year="1"}}
     \if@filesw
     \immediate\write\@auxout{\string\citation{apsrev41Control}}%
    \fi

\begin{document}

\title{The Effect of Geometry, Spin and Orbital Optimization in Achieving Accurate, Fully-Correlated Results for Iron-Sulfur Cubanes}
\date{\today}
\author{Carlos~Mejuto-Zaera}
\thanks{carlos\_mejutozaera@berkeley.edu}
\affiliation{University of California, Berkeley, California 94720, United States}
\affiliation{Computational Research Division, Lawrence Berkeley National Laboratory, Berkeley, CA 94720, USA}
\author{Demeter Tzeli}
\affiliation{Laboratory of Physical Chemistry, Department of Chemistry, National and Kapodistrian University of Athens, Panepistimiopolis Zografou, Athens 15784, Greece}
\affiliation{Theoretical and Physical Chemistry Institute,
National Hellenic Research Foundation, Vas. Constantinou 48, Athens 11635, Greece}
\author{David Williams-Young}
\affiliation{Computational Research Division, Lawrence Berkeley National Laboratory, Berkeley, CA 94720, USA}
\author{Norm~M.~Tubman}
\thanks{norman.m.tubman@nasa.gov}
\affiliation{Quantum Artificial Intelligence Lab. (QuAIL), Exploration Technology Directorate,
NASA Ames Research Center, Moffett Field, CA 94035, USA}
\author{Mikul\'{a}\v{s} Matou\v{s}ek}
\affiliation{J. Heyrovský Institute of Physical Chemistry, Academy of Sciences of the Czech Republic, v.v.i., Dolejškova 3, 18223 Prague 8, Czech Republic}
\author{Jiri Brabec}
\affiliation{J. Heyrovský Institute of Physical Chemistry, Academy of Sciences of the Czech Republic, v.v.i., Dolejškova 3, 18223 Prague 8, Czech Republic}
\author{Libor Veis}
\thanks{libor.veis@jh-inst.cas.cz}
\affiliation{J. Heyrovský Institute of Physical Chemistry, Academy of Sciences of the Czech Republic, v.v.i., Dolejškova 3, 18223 Prague 8, Czech Republic}
\author{Sotiris S. Xantheas}
\email{sotiris.xantheas@pnnl.gov}
\affiliation
{Advanced Computing, Mathematics and Data Division, Pacific Northwest National Laboratory, 902 Battelle Boulevard, P.O. Box 999, MS K1-83, Richland, WA 99352, USA}
\affiliation{Department of Chemistry, University of Washington, Seattle, WA 98185, USA}
\author{Wibe~A.~de~Jong}
\thanks{WAdeJong@lbl.gov}
\affiliation{Computational Research Division, Lawrence Berkeley National Laboratory, Berkeley, CA 94720, USA}

\begin{abstract}
Iron-sulfur clusters comprise an important functional motif of the catalytic centers of biological systems, capable of enabling important chemical transformations at ambient conditions. This remarkable capability derives from a notoriously complex electronic structure that is characterized by a high density of states that is sensitive to geometric changes. The spectral sensitivity to subtle geometric changes has received little attention from fully-correlated calculations, owing partly to the exceptional computational complexity for treating these large and correlated systems accurately. To provide insight into this aspect, we report the first Complete Active Space Self Consistent Field (CASSCF) calculations for different geometries of cubane-based clusters using two complementary, fully-correlated solvers: spin-pure Adaptive Sampling Configuration Interaction (ASCI) and Density Matrix Renormalization Group (DMRG). We find that the previously established picture of a double-exchange driven magnetic structure, with minute energy gaps ($< 1$~mHa) between consecutive spin states, has a weak dependence on the underlying geometry. However, the spin gap between the lowest singlet and the highest spin states is strongly geometry dependent, changing by an order of magnitude upon slight deformations that are still within biologically relevant parameters. The CASSCF orbital optimization procedure, using active spaces as large as 86 electrons in 52 orbitals, was found to reduce this gap by a factor of two compared to typical mean-field orbital approaches. Our results clearly demonstrate the need for performing highly correlated calculations to unveil the challenging electronic structure of these complex catalytic centers.
\end{abstract}

\maketitle

\section{Introduction}
\label{sec:intro}

Iron–sulfur clusters are ubiquitous. They are involved in many biological systems operating as active centers of proteins in essential life-sustaining processes such as photosynthesis, respiration, and nitrogen fixation.~\cite{Spiro1982,Holm1996,Kessler2005} They are involved in electron transfer processes,~\cite{Jang2000,Brzoska2006} substrate activation and binding,~\cite{Einsle2002,Doukov2002} catalytic reactions,~\cite{Berkovitch2004,Munck2008} DNA repair,~\cite{Lukianova2005} signal transductions,~\cite{Kiley2003} iron/sulfur storage,~\cite{Beinert1997} regulation of gene expression,~\cite{Runyen-Janecky2008} and enzyme activity.~\cite{Wofford2015} Additionally, they are significant in industrial catalysis.~\cite{Rees2003,Nurmaganbetova2001} The key for their remarkable reactivity is their low-lying, dense electronic state manifold. Of particular importance is the understanding of the intrinsic electronic structure of the Fe-S clusters as well as its modification due to their surroundings as prerequisites in order to interpret their functionality and properties. As the result of the continuing interest in these iron-sulfur systems, several previous investigations have been reported.

There have been many previous computational studies reported for iron-sulfur clusters employing the Broken Symmetry analysis~\cite{Noodleman1988} of spin coupling  and especially the commonly used BS-DFT methodology, see for instance.~\cite{Sigfridsson2001,Niu2009,Dance2015,Bergeler2013,Carvalho2014} In general, this approach works quite well for the prediction of the geometry for molecular clusters involving multiple transition metals. However, it describes a weighted average over the (multiplet) states and as such it is not appropriate enough for the efficient calculation of the correlation energy of these multi-reference systems. Additionally, it depends on the density functional used.~\cite{Neese2009}  Note that previous calculations on the [Fe$_2$S$_2$(SCH$_3$)$_2$]$^{-2}$ and [Fe$_4$S$_4$(SCH$_3$)$_4$]$^{-2}$  clusters indicated that both clusters have an unusually dense spectrum which is different from the predictions of the Heisenberg double-exchange model.~\cite{Sharma2014}

While both tri-nuclear [Fe$_3$S$_4$] and tetra-nuclear [Fe$_4$S$_4$] clusters are found in proteins such as ferredoxins and are both regarded as electron transfer sites in a variety of bacteria,~\cite{Schipke1999,Kissinger1991} much attention has been mainly given to the [Fe$_4$S$_4$] clusters by applying mainly the BS-DFT methodology. To the best of our knowledge, there are only two previous theoretical studies where multi-reference methodologies, such as Density Matrix Renormalization Group (DMRG)~\cite{Sharma2014} and Coupled Cluster Valence Bond (CCVB),~\cite{Small2018}  have been applied to just the [Fe$_4$S$_4$(SCH$_3$)$_4$]$^{-2}$ cluster. It should also be noted multi-reference methods, such as MC-PDFT, CASPT2/RASPT2 and NEVPT2, have been used for bimetallic Fe-S clusters.~\cite{Presti2019}

In this study we report the results for the [Fe$_4$S$_4$(SCH$_3$)$_4$]$^{-2}$ cluster using the Adaptive Sampling Configuration Interaction (ASCI) and DMRG methodologies. These are variational and complementary methods to treat strong correlation in many-body systems as in the present case. The purpose of using the above methodologies is to capture both the static and dynamic correlation within the active space, while the ASCI plus second order perturbation (ASCI+PT2) extrapolated results are used to estimate the Full Configuration Interaction (FCI) limit.

\section{Methods}
\label{sec:methods}

To study the low energy eigenstates of iron-sulfur clusters, we employ the ASCI~\cite{Tubman2016,Tubman2018a,Tubman2020} and DMRG~\cite{White1992,Schollwock2011,Chan2011} approaches, both for ground state calculations and as approximate solvers in  CASSCF~\cite{Helgaker2000,Wouters2014,Levine2020,Brabec2020} orbital optimizations. ASCI and DMRG are complementary methods to treat strong correlation in many-body systems, based on different heuristics: the former finds the most relevant Slater determinants for a truncated ground state description exploiting perturbative estimates iteratively, whereas the latter leverages the simple orbital entanglement structure in ground state wave functions to determine a compact Matrix Product State (MPS) wave function. \cite{Schollwock2011}

Here, we briefly outline the two methods, and refer the reader to the relevant literature and the Electronic Supporting Information (ESI) for further details. Additionally, we present a new flavor of ASCI to target pure spin states, based on organizing the Slater determinants in Configuration State Function (CSF) families. This approach, which we label SP-ASCI, is necessary to avoid spin contamination in the truncated wave functions for the iron-sulfur clusters.

\subsection{ASCI and ASCI-SCF}

The ASCI approach relies on an exceptionally efficient Selected Configuration Interaction (SCI) protocol to describe ground states. Using an iterative approach based on perturbative estimations,~\cite{Huron1973,Evangelisti1983,Illas1991} ASCI can identify the determinants in the Hilbert space that have large coefficients in the ground state wave function. Truncating the full Hilbert space to this determinant subset and subsequently projecting the Hamiltonian operator, results in highly compact approximate wave functions, which can nonetheless capture the static correlation of many-body systems accurately. This typically requires an active space formulation, and hence ASCI is successful in describing multi-reference systems with a limited number (i.e. less than 50) of correlated orbitals. Dynamical correlation within the active space can then be recovered perturbatively,~\cite{Tubman2018a} and ASCI has been shown to provide near FCI accuracy for the ground state energies and spectral functions for a wide variety of challenging, strongly correlated molecular and extended systems.~\cite{Tubman2016,Tubman2018a,Tubman2018b,Mejuto2019,Hait2019,Mejuto2020b,Tubman2020,Eriksen2020,Levine2020}  As is usual in SCI approaches, the orbital basis chosen to define the Hamiltonian has a critical effect on the convergence of ASCI, and simple choices such as the natural orbital basis~\cite{Lowdin1955,Lowdin1956a,Lowdin1956b,Davidson1972} do not always assure rapid convergence. It is for this purpose that the more sophisticated CASSCF orbital optimization can provide a decisive advantage, since it determines the variationally optimal orbital basis for a multi-reference wave function. Using ASCI in conjunction with CASSCF has been shown to enable the study of large active spaces in transition metal systems,~\cite{Levine2020,Zhao2020} and for this reason we have chosen to employ this method for the study of the iron-sulfur clusters.

\subsection{Spin Pure ASCI Mimicking CSFs}

While ASCI has been shown to successfully propose highly accurate ground state truncations, being a SCI approach it is susceptible to breaking symmetries. This happens when the generator of the symmetry $O$ and the Hamiltonian $H$ do not commute after being projected to the ASCI truncation. For (non-relativistic) systems with strong magnetic character, such as the iron-sulfur clusters in this work, the breaking of spin symmetry (i.e. $O = S^2_{tot}$) can become a major computational problem, and we did indeed observe a large degree of spin contamination using ASCI even for modest active spaces in these systems. This difficulty arises from the fact that the iterative search for an optimal truncation in ASCI is formulated in terms of single Slater determinants, and these are not generally eigenstates of the total spin operator~\cite{Helgaker2000}. Thus, we resolve the spin-contamination problem by building the ASCI truncation in terms of groups of determinants spanning CSF families. These correspond to all possible determinants with a specified occupation scheme, defined by which orbitals are empty, singly and doubly occupied. In this way, the ASCI truncation is guaranteed to preserve spin symmetry. We further select the eigenstate with the smallest possible spin quantum number by adding a spin penalty term $\lambda S^2_{tot}$ to the Hamiltonian during the energy calculation. This novel approach, which we label spin pure ASCI (SP-ASCI), enables treating targeted spin states in strongly correlated systems accurately, as we show in the results section. We refer to the ESI for details on the implementation and a discussion with related existing methods to resolve spin contamination in SCI-related electronic structure algorithms.~\cite{Fales2017,Applencourt2018,Li2020}

\subsection{DMRG and DMRG-SCF}

DMRG is a very powerful approach suitable for the treatment of strongly correlated systems that was
originally 
developed 
in solid state physics. \cite{White1992,White-1992b,White-1993} It has been established as one of the reference methods for the electronic structure calculations of strongly correlated molecules requiring very large active spaces.\cite{Chan2011, wouters_review,legeza_review,yanai_review, reiher_perspective} Complexes with multiple transition metal centers are, due to the large quasi-degeneracy of \textit{d} shells, typical examples of such species and belong to the most advanced quantum chemical applications of DMRG.\cite{kurashige_2013, Sharma2014, Li2019, chan_new, Brabec2020}

The DMRG method is a variational procedure, which optimizes the wave function in the form of MPS. \cite{Schollwock2011} The practical version of DMRG is the two-site algorithm, which provides the wave function in the two-site MPS form

\begin{eqnarray}
  \label{eq:MPS_2site}
  | \Psi_\text{MPS} \rangle = \sum_{\{\alpha\}} \mathbf{A}^{\alpha_1} \mathbf{A}^{\alpha_2} \cdots \mathbf{W}^{\alpha_i \alpha_{i+1}} \cdots \mathbf{A}^{\alpha_n}| \alpha_1 \alpha_2 \cdots \alpha_n \rangle, \nonumber \\
\end{eqnarray}

where $\alpha_i \in \{ | 0 \rangle, | \downarrow \rangle, | \uparrow \rangle, | \downarrow \uparrow \rangle \}$ 
for a given pair of adjacent indices $[i, (i+1)]$, $\mathbf{W}$ is a four index tensor, which corresponds to the eigenfunction of the electronic Hamiltonian expanded in the tensor product space of four
tensor spaces defined on an ordered orbital chain, so called \textit{left block} ($M_l$ dimensional tensor space), \textit{left site} (four dimensional tensor space of $i^{\text{th}}$ orbital), \textit{right site} (four dimensional tensor space of $(i+1)^{\text{th}}$ orbital), and \textit{right block} ($M_r$ dimensional tensor space). 
The MPS matrices $\mathbf{A}$ are obtained by successive application of the singular value decomposition (SVD) with truncation on the $\mathbf{W}$'s and iterative optimization by 
going through the ordered orbital chain from \textit{left} to \textit{right} and then sweeping back and forth. \cite{legeza_review} The maximum dimension of MPS matrices which is required for a given accuracy, so called bond dimension
$[M_{\text{max}} = \text{max}(M_l, M_r)]$,
can be regarded as a function of the level of entanglement in the studied system. \cite{legeza_2003b} 
Among others, $M_{\text{max}}$ strongly depends on the order of orbitals along the one-dimensional chain \cite{legeza_2003a, moritz_2005} as well as their type.\cite{fertitta_2014, Krumnow2016, amaya_2015}

Similarly to ASCI, DMRG can replace the exact diagonalization in the CASSCF procedure, which leads to the formulation of the method usually denoted as DMRG-SCF. \cite{Zgid-2008c, Ghosh-2008} Since different elements of 2-RDMs are collected at different iterations of the DMRG sweep,\cite{Zgid-2008b}
the one-site DMRG algorithm has to be used for the final computations of the 2-RDMs to assure the same accuracy of all their elements. \cite{Zgid-2008c}

As was mentioned above, the studied systems are prone to spin contamination. There exist spin-adapted formulations of quantum chemical DMRG, \cite{Sharma-2012a, Wouters-2014a, keller_2016} however since the spin-adapted version of the MOLMPS program, \cite{Brabec2020} which was used in the current study, is under development, we used an approach similar to the one described in the previous section, in particular we penalized higher spin states of the given spin projection by an additional term added to the Hamiltonian ($\lambda S^2_{tot}$).


\section{Methodological and System Details}
\label{sec:GeomsAndDetails}

\begin{figure}
\includegraphics[width=0.4\textwidth]{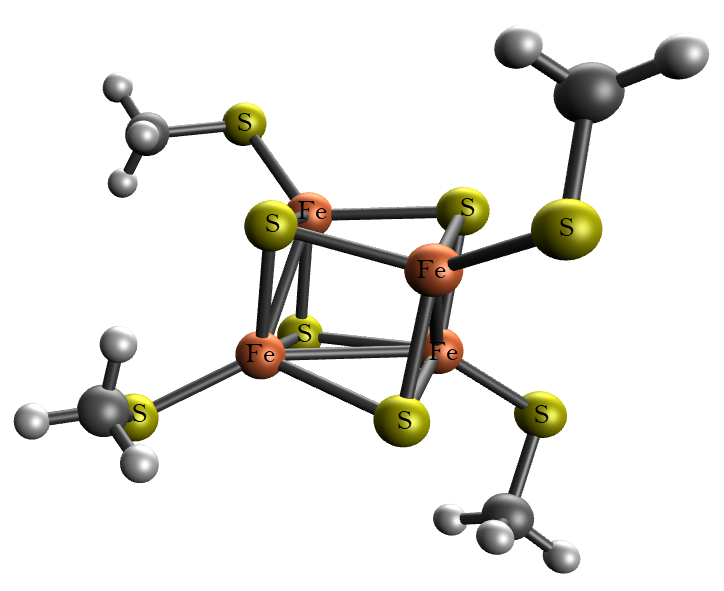}
\caption{\label{fig:geoms} Sample geometry 2A for the [Fe$_4$S$_4$(SMe)$_4$]$^{-2}$ clusters.}
\end{figure}

In this study we consider the [Fe$_4$S$_4$(SMe)$_4$]$^{-2}$ clusters, with  Me = CH$_3$, which correspond to a family of comparatively simple model systems of experimental and theoretical interest, since they serve as approximate structural motifs of the complex iron based catalytic centers in biologically relevant enzymes.~\cite{Zhang1996,Nakajima1997,Wang2003,Waters2007,Sharma2014,Derosha2019} Formally, the -2 charge corresponds to an average  oxidation number of $+2.5$, which can be interpreted as two pairs of Fe(II) and Fe(III) atoms. To study the effect of cluster geometry on the energy and spin ordering, we performed broken-symmetry (BS) DFT geometry optimizations for different spin states of this cluster, using the HTPPSh functional and a mixed aug-cc-pVDZ (Fe, S) and cc-pVDZ (H,C) basis set, which we denote as (aug)-cc-pVDZ. This basis set has 568 basis functions. We determined 3 geometries, labeled 2A, 2B and 2C. We show a sample (2A) geometry in Fig.~\ref{fig:geoms}, and refer to the Supporting Information~\cite{supp} for the .xyz files containing the Cartesian coordinates. In Table~\ref{tab:key} we provide the summary of the cluster labels and spin states, together with the BS-DFT energies. We also used the geometry in Ref.~\cite{Sharma2014}, labeled as 2R, for comparison purposes. In the ESI~\cite{supp}, we further include the ROHF and CASSCF energies for 5 geometries of the [Fe$_3$S$_4$(SMe)$_3$]$^{-2}$ clusters, which we used as test systems to benchmark our algorithms, as well as similar results for the 2R geometry with the def2-SVP basis set, to offer a direct comparison with the results reported in Ref. ~\cite{Sharma2014}.


\begin{table}[hbt!]
  \begin{center}
    \begin{tabular}{|l|c|c||c|} 
    \hline
    Cluster & $2S + 1$ & BS-DFT & HS-ROHF ($2S+1$)  \\
    \hline
    2A (Fe$_4$) &  19 & -8401.048191 & -8389.083540 (19) \\
    2B (Fe$_4$) & 1  & -8401.092653 & -8389.026178 (19) \\
    2C (Fe$_4$) &  1  & -8401.092522 & -8389.024431 (19) \\
    \hline
    \hline
    2R\cite{Sharma2014} (Fe$_4$) & -- & -- & -8388.979087 (19) \\
    \hline
    \end{tabular}
  \end{center}
  \caption{Summary table relating the [Fe$_4$S$_4$(SMe)$_4$]$^{-2}$ cluster labels to their geometries. Each geometry was obtained by performing a BS-DFT optimization with different target spin states. We considered three geometries. The BS-DFT calculations used a mixed aug-cc-pVDZ (Fe, S) cc-pVDZ (C, H) basis set and the TPSSh functional. We also include the geometry in Ref.~\cite{Sharma2014} in our study as reference. We further list the ROHF energies for the high spin (HS) states in each cluster geometry, which we localized and used as the starting point for the CASSCF orbital optimizations.}
  \label{tab:key}
\end{table}

The cluster geometry plays a crucial role in the catalytic activity of iron-sulfur clusters, particularly considering the high tunability of the protein environment in which they are often embedded. For the cubane systems considered in this work, the relevant parameters describing the Fe-S ``cube'', i.e., the Fe-S bond lengths, the Fe--Fe distances and the Fe-S-Fe angles, are equivalent to averages in crystallographic data for similar compounds.~\cite{Averill1973,Que1974} These geometries are significantly different from an ideal cube. The average Fe-S-Fe angles are $\sim75^\circ$, in very good agreement with the experimental values where the corresponding average angle is $73.81^\circ$~\cite{Averill1973}, but significantly smaller than a right angle that is present in an ideal cube. Similarly, in a perfect cube, the average Fe--Fe distance would be exactly $\sqrt{2}$ times the average Fe-S bond length. However, our geometries show deviations of $12-16\%$ from this ideal relation, consistent with the deviations of 15$\%$ in the experimental geometry. Besides the departure from the ideal cubic geometry, these clusters present a further asymmetry: within the Fe--Fe distances in a given cluster, there are always two Fe-pairs with shorter distance than all other possible combinations. For geometries 2B and 2R, this difference is only slight (the shorter Fe--Fe distances are $\sim3\%$ smaller), but geometry 2A has a more pronounced asymmetry with the shorter Fe--Fe distance being $\sim 11\%$ smaller. As our results show, this subtle difference in the geometry has huge consequences to the electronic structure: it reduces the largest spin gap, i.e. the energy difference between the most stable singlet and the $2S+1=19$ configurations, by one order of magnitude! This is a remarkable manifestation of the electronic tunability of these catalytic centers as a result of small variations in the cube's geometry.

Experimentally, the geometries of [Fe$_4$S$_4$(SCH$_2$Ph)$_4$]$^{2-}$ and [Fe$_4$S$_4$(SPh)$_4$]$^{2-}$ have been measured; the average Fe-Fe distances are 2.747~{\AA} ~\cite{Averill1973} and 2.736~{\AA}, ~\cite{Que1974}  respectively, while the average Fe-S distance is 2.286~{\AA} for both anions. Comparing our calculated average Fe-Fe and Fe-S distances with these available crystallographic data, we found that our average calculated Fe-Fe distances of the 2A-2C structures are about 0.5-8 $\%$ elongated, while our Fe-S distances are elongated about 2$\%$-4$\%$ with respect to the experimental data. These deviations are reasonable, given that the experimental data correspond to derivatives of our calculated [Fe$_4$S$_4$(SMe)$_4$]$^{2-}$ units in the solid state, and hence we are confident that our geometries are representative of the actual Fe-S clusters in catalytic centers. 

\section{Results}
\label{sec:results}

For each cluster geometry [Fe$_4$S$_4$(SMe)$_4$]$^{-2}$ and spin state, we optimized its orbitals using the following protocol: First, we performed a Restricted Open-Shell Hartree Fock (ROHF) calculation on a high-spin (HS) state, nameley $2S+1 = 19$. We report the HS-ROHF energies, obtained with the (aug)-cc-pVDZ basis set, in the last column of Tab.~\ref{tab:key}.  Already at the mean-field level, we can observe a huge effect on the energy due to slight geometry variations. In particular, we note that with the mixed (aug)-cc-pVDZ  basis set, the reference geometry 2R is not optimal for the oxidation state considered here. Further details regarding the ROHF starting points can be found in the ESI~\cite{supp}.
Those ROHF orbitals then serve as starting points for CASSCF orbital optimizations. 
We used a (54e, 36o) active space, including Fe-\textit{d}, bridge S-\textit{p} orbitals, and the ligand S-\textit{p} orbitals that point into the corresponding Fe atom. To simplify the active space identification, we performed a Pipek-Mezey~\cite{Pipek1989} orbital localization of the core and valence orbitals separately, as has been done in previous studies of these systems.~\cite{Sharma2014} 
Despite the fact that neither the ROHF, nor the ASCI and DMRG energies are invariant under such localization (due to mixing of open- and closed-shells), 
it is advantageous for the post-HF processing since it allows us to choose chemically motivated active spaces easily, it improves the CASSCF convergence, and further simplifies the interpretation of correlation functions in the results section. 
We used SP-ASCI as the CAS solver approximation in these active spaces. Following the suggestions in Ref.~\cite{Levine2020}, we performed a step-wise orbital optimization, systematically increasing the number of determinants in the ASCI solver, starting with $10^5$ determinants, and increasing as: $2.5\cdot10^5$, $10^6$, $2\cdot10^6$, $5\cdot10^6$. In terms of CSF families, this sequence corresponds for the (52e, 36o) active space to approximately 4, 20, 50, 100, 230 CSF families respectively. Although the first two optimizations contain a small number of CSF families, we have observed that we reach lower energies by starting there, as opposed to starting the CASSCF optimization from the ROHF orbitals directly with a $10^6$ ASCI-SCF calculation. This may point to some effective pre-optimization of the inactive orbitals.

These CASSCF optimizations are highly complex, prone to falling into local minima. This becomes particularly apparent when comparing the CASSCF energies for different spin states of the same geometry, which often would show gaps of several mHa, in extreme cases even $>10$ mHa. In these cases, following the suggestions in Ref.~\cite{Levine2020}, we attempt to escape the local minima of, for example, spin state $S$ by restarting its CASSCF with the optimized orbitals of spin state $S \pm 2$. Alternatively, within the SP-ASCI framework, we can restart the CASSCF with orbitals obtained from a low determinant ASCI-SCF without imposing the total-spin conservation symmetry. We continue this process until the CASSCF energy converges to within 1~mHa. It is important to note that this strategy does not guarantee reaching the global minimum during the CASSCF optimization, and instead the results correspond just to stable local minima. This is unavoidable in complex non-linear optimizations as the ones performed in this work, involving 584 basis functions and large active spaces. Still, obtaining physical magnitudes, such as spin gaps and correlation functions, in good agreement with the theoretical and experimental literature gives us confidence that our results are representative of the actual chemistry of these clusters.

To obtain an estimate of the FCI energy within the (54e, 36o) active space, we performed ASCI+PT2 extrapolations~\cite{Tubman2018a} with the optimized CASSCF orbitals. For this, we computed ASCI+PT2 energies for truncations of $5\cdot10^5$, $10^6$, $2\cdot10^6$, $5\cdot10^6$ determinants, and then fit the ASCI+PT2 energy vs the PT2 correction as a straight line. The value for the fitted $y$ intercept corresponds to our best estimate for the energy with no perturbtative correction, i.e. the FCI limit. We report the uncertainty of the $y$ intercept as a measure of the systematic error of the extrapolation. Notably, this extrapolation is less straightforward for the Fe-S cubanes than for smaller systems previously studied with ASCI, since the ASCI+PT2 energies do not follow a perfect linear dependency as a function of the PT2 correction. Still, this offers a viable estimate of the missing dynamical correlation in the ASCI wave function.

Given the inherent complexity of the electronic structure in Fe-S clusters, we subsequently investigate them with DMRG, which as discussed above is based on a different heuristic than ASCI. We use the CASSCF orbitals optimized by ASCI as starting points for further CASSCF orbital optimization, this time using the DMRG approach as a solver and bond dimensions $M=2000$. In the case of DMRG-SCF, active-active orbital rotations were not considered. Once this subsequent approximation is converged, we performed 
accurate DMRG calculations with the dynamical block state selection (DBSS) \cite{legeza_2003a} and predefined truncation error TRE = $10^{-6}$ (unless otherwise stated, depending on the structure and spin state, this corresponds up to $M=16000$).

\begin{table}[t!]
  \begin{center}
    \begin{tabular}{|l|c|c|c|} 
    \hline
    Cluster & $2S + 1$ & ASCI+PT2 extrapol. & DMRG \\
    Geom. &          &                    & (TRE $=5\cdot10^{-6}$) \\
    \hline
    2A  & 5 & -9.020195 $\pm$ 0.002354 & -9.011678 \\
      & 3 & -8.958029 $\pm$ 0.007356  & -9.011427 \\
      & 1 & -8.964512 $\pm$ 0.010117 & -9.012364 \\
    \hline
    2B  & 5 & -8.897534 $\pm$ 0.038871 & -8.989614 \\
     & 3 & -8.944783 $\pm$ 0.009405 & -8.990533 \\
     & 1 & -8.905407 $\pm$ 0.016850 & -8.991705 \\
    \hline
    2C  & 5 & -8.868532 $\pm$ 0.045080 & -8.985393 \\
     & 3 & -8.901021 $\pm$ 0.011785 & -8.986240\\
    & 1 & -8.875057 $\pm$ 0.013729  &  -8.987101 \\
    \hline 
    2R  & 5 & -8.877203 $\pm$ 0.014616 & -8.928211\\
     & 3 & -8.882806 $\pm$ 0.016360 & -8.929151 \\
     & 1 & -8.848366 $\pm$ 0.016622 &  -8.930184 \\
    \hline
    \end{tabular}
  \end{center}
  \caption{Extrapolated ASCI and DMRG energies ($E-8380.0$ Ha) for the (54e, 36o) active space of [Fe$_4$S$_4$(SMe)$_4$]$^{-2}$, using the high spin ($2S + 1 = 19$) ROHF orbitals, localized with Pipek-Mezey, with the (aug)-cc-pVDZ basis. The ASCI-PT2 extrapolations to the FCI limit use a linear extrapolation from calculations with $5\cdot10^5$, $1\cdot10^6$, $2\cdot10^6$, $5\cdot10^6$ determinants. The error bars correspond to the standard deviation of the linear fit, and are thus just a measure of the extrapolation error alone. The DMRG energies for the 2R geometry used TRE $=10^{-5}$.  
  }
  \label{tab:MB_ROHF_Fe4}
\end{table}

Finally, we note that we performed additional calculations on some of the clusters to better illustrate our conclusions. For instance, we performed ASCI-PT2 and DMRG calculations on the unoptimized, although localized, ROHF orbitals, in order to investigate the effect of the CASSCF optimization, as well as ASCI calculations on the high-spin state using the singlet CASSCF optimized orbitals to estimate total spin gaps. Furthermore, we tested different active spaces for the 2A cluster, and include a brief note on basis set choice in the ESI~\cite{supp}. We will introduce the details of these additional calculations whenever pertinent.

\subsection{ASCI and DMRG with Hartree-Fock orbitals}

In order to assess the effect of the orbital optimization on the physical description of the iron-sulfur clusters, we first compute the energies for the low lying spin states with the unoptimized HS-ROHF orbitals, localized via the Pipek-Mezey scheme. We report the variational ASCI and DMRG energies in the (54e, 36o) active space for the Fe$_4$ clusters in Tab.~\ref{tab:MB_ROHF_Fe4}. To provide a comparison with the existing literature, we report the energies of the low lying spin states for the 2R geometry using the def2-SVP basis set in the ESI.

We observe that the ASCI and DMRG energies in Tab.~\ref{tab:MB_ROHF_Fe4} follow the same hierarchy as the mean-field energies in Tab.~\ref{tab:key}, with 2A being the most stable geometry and 2R the least stable one. The DMRG energy gaps between consecutive spin states are nonetheless comparable between geometries, typically $\sim1$~mHa. Notice that 
the ASCI and DMRG energies in Tab.~\ref{tab:MB_ROHF_Fe4} are above the canonical ROHF energies in Tab.~\ref{tab:key}, which is due to the aforementioned localization of ROHF orbitals mixing open- and closed-shells.

We note that the ASCI+PT2 extrapolated energies in Tab.~\ref{tab:MB_ROHF_Fe4} show large gaps of tens of mHa between the different spin states, as well as comparably large extrapolation errors. These gaps are not consistent with either previous results in the literature,~\cite{Sharma2014} or with our DMRG calculations in the right most column of Tab.~\ref{tab:MB_ROHF_Fe4}, or indeed with the trends observed for the Fe$_3$ clusters in the ESI~\cite{supp}. Furthermore, the PT2 extrapolations in these systems are rather unreliable, since the perturbative corrections are quite large ($>$ 60 mHa), and the convergence behavior is far from linear in several of the spin states and used geometries. Indeed, for the 2A, 2B and 2C geometries we dismissed, as a clear outlier, the $5\cdot10^5$ ASCI calculation in the extrapolation since it shows poor convergence towards the FCI limit. These issues are rather symptomatic of ASCI converging very slowly in the localized ROHF orbital basis. For smaller molecular systems, rotations to an approximate natural orbital basis drastically improve the ASCI convergence. Unfortunately, these cubane clusters have a too pronounced multi-reference character in the localized ROHF basis, such that even at 5 million determinants the 1-RDM is not representative for the true ground state, and thus the corresponding natural orbital rotation does not resolve the convergence problem. Instead, a more sophisticated single-particle rotation is needed for ASCI to provide reliable results in these exceptionally complex systems, and thus we turn our attention to CASSCF orbital optimization.

\subsection{CASSCF with ASCI and DMRG}

\begin{table*}[hbt!]
  \begin{center}
    \begin{tabular}{|l|c|c|c|c|c|} 
    \hline
    Cluster & $2S + 1$ & CASSCF & ASCI+PT2 extrapol. &  DMRG-SCF & DMRG \\
    Geom. &          & SP-ASCI-SCF & SP-ASCI &  ($M=2000$) & (TRE = 10$^{-6}$) \\
    \hline
    2A & 5 & -9.150451 & -9.155331 $\pm$ 0.000981  & -9.154563 & -9.155436 \\
     & 3 & -9.150856 & -9.155357 $\pm$ 0.000843 & -9.155611 & -9.155571 \\
     & 1 & -9.150950 & -9.155423 $\pm$ 0.000681 & -9.155072 & -9.155582 \\
    \hline
    2B & 5 & -9.117029 & -9.123016 $\pm$ 0.001326 & -9.124550 & -9.126843 \\
     & 3 & -9.117290 & -9.123942 $\pm$ 0.000914 & -9.124132 & -9.126312 \\
     & 1 & -9.117792 & -9.123909 $\pm$ 0.001061 & -9.122910 & -9.126381 \\
    \hline
    2C & 5 & -9.089257 & -9.095255 $\pm$ 0.000966 & -9.089034 & -9.092818 \\
     & 3 & -9.088027 & -9.092412 $\pm$ 0.001317 & -9.088379 & -9.091081 \\
     & 1 & -9.090888 & -9.096716 $\pm$ 0.001695 & -9.088452 & -9.091410 \\
    \hline 
    2R & 5 & -9.063877 & -9.071370 $\pm$ 0.000992  & -9.073519 & -9.074443 \\
     & 3 & -9.062709 & -9.068967 $\pm$ 0.000990 & -9.071405 & -9.072118 \\
     & 1 & -9.063875 & -9.070530 $\pm$ 0.000655 & -9.069628 & -9.072103 \\
    \hline
    \end{tabular}
  \end{center}
  \caption{CASSCF and extrapolated energies ($E - 8380.0$ Ha) using SP-ASCI and DMRG for the (54e, 36o) active space of [Fe$_4$S$_4$(SMe)$_4$]$^{-2}$, starting from the high spin ($2S + 1 = 19$) ROHF orbitals, localized with the Pipek-Mezey scheme, with the (aug)-cc-pVDZ basis. For the CASSCF energies with ASCI, the results correspond to calculations with $5\cdot10^6$ determinants. These are the final steps of a series of SP-ASCI-SCF calculations starting at $1\cdot10^5$ determinants, and progressively increasing the number of determinants to improve the orbitals sequentially. The ASCI+PT2 extrapolated results are estimating the FCI limit using a linear extrapolation from calculations with $5\cdot10^5$, $1\cdot10^6$, $2\cdot10^6$, $5\cdot10^6$ determinants, starting from the orbitals obtained from the SP-ASCI-SCF with $5\cdot10^6$ determinants. The error bars correspond to the standard deviation of the linear fit, and are thus just a measure of the extrapolation error alone. The extrapolation for the 2C singlet state was performed considering only the last with 3 calculations.}
  \label{tab:MCSCFFe4}
\end{table*}

The CASSCF energies for the low lying spin states for the four cluster geometries, employing SP-ASCI and DMRG as solvers with a (52e, 36o) active space, are summarized in Tab.~\ref{tab:MCSCFFe4}. 
In the case of ASCI we report extrapolated energies to the FCI limit, complementing ASCI with second order perturbation theory~\cite{Tubman2018a}. 
For accurate DMRG calculations, we have used the tight truncation error criterion TRE = $10^{-6}$.

\begin{figure*}
    \centering
    \includegraphics[width=\textwidth]{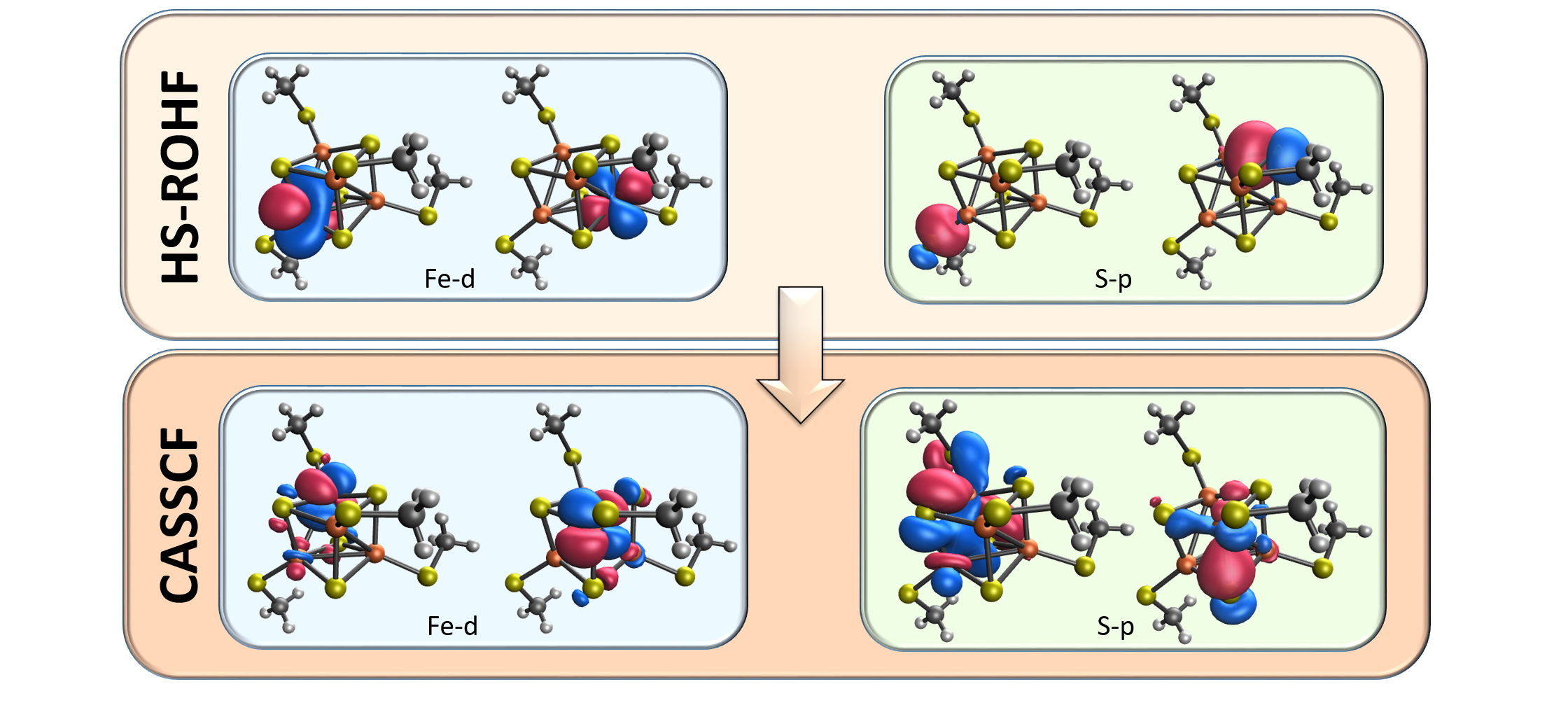}
    \caption{Sample orbitals before (HS-ROHF) and after the CASSCF rotation, for the 2A [Fe$_4$S$_4$(SMe)$_4$]$^{-2}$ cluster geometry, spin state $2S+1=1$. The CASSCF results correspond to the final optimization, using the SP-ASCI solver with $5\cdot10^6$ determinants. For the localized HS-ROHF, we show two orbitals from the Fe-\textit{d} set, and 2 orbitals from the S-\textit{p} set. For the CASSCF calculation we select two orbitals each of the same sections of the active space, which present significant changes with respect to the localized HS-ROHF orbitals.}
    \label{fig:2A_sing_orbrot}
\end{figure*}

The different geometries show minute energy gaps between subsequent spin states, listed in Tab.~\ref{tab:MCSCFFe4}. We note the excellent agreement between ASCI and DMRG. The latter having been validated as CASSCF solver against FCI, shows that ASCI also provides an accurate CAS approximation to use in the orbital optimization.\footnote{It is important to remember that this does not mean that we have reached a global minimum in the CASSCF optimization, but that the DMRG-SCF optimization subsequent to ASCI-SCF did not find a better set of orbitals in these cases. We note that this is still a possible outcome, and in particular we show an example of this situation in the ESI for the def2-SVP basis set calculations in the 2R geometry.} Furthermore, the CASSCF optimized orbitals resolve the convergence issues that ASCI presents in the ROHF orbital basis, resulting in more reliable perturbative extrapolations, with errors of $\sim$1~mHa. For all geometries, we observe the singlet state to be the most stable at the variational CASSCF level, with the exception of the 2R geometry, in which the singlet and quintet are essentially degenerate. In general, the observed energy gaps are approximately 1~mHa or smaller. Upon extrapolation to the full CI limit with ASCI+PT2, the gaps remain small, and some degree of spin reordering is apparent. However, since in this case the gaps are of the order of magnitude of our extrapolation error, it is not possible to make any definitive statement about the actual spin orderings in the FCI limit.

\begin{table}[hbt!]
  \begin{center}
    \begin{tabular}{|l|c|c|c|c|c|} 
    \hline
    Geometry & $2S + 1 = 1$ & $2S + 1 = 19$ & Spin Gap [mHa] \\
    \hline
    2A & -9.150950 & -9.149944 & -1.006 (-3.471) \\
    2B & -9.117792 & -9.108935 & -8.857 (-17.879) \\
    2C & -9.090888 & -9.081854 & -9.034 (-16.986) \\
    2R & -9.063875 & -9.054035 & -9.840 (-18.490) \\
    \hline
    \end{tabular}
  \end{center}
  \caption{ASCI energies ($E - 8380.0$ Ha) using SP-ASCI as solver for the different [Fe$_4$S$_4$(SMe)$_4$]$^{-2}$ cluster geometries in the $2S+1 = 1$ and $2S+1 = 19$ spin states for the (54e, 36o) active space with (aug)-cc-pVDZ basis. The energies are computed using the CASSCF orbitals optimized for the singlet state. The last column reports the spin gap in mHa, computed as ${E_0^{2S+1 = 1} - E_0^{2S+1=19}}$. The numbers in parenthesis are the spin gaps computed in the localized ROHF basis using DMRG.}
  \label{tab:SpinGapGeoms}
\end{table}

It is possible to make stronger claims about the largest spin gap in the systems, i.e. the gap between the lowest ($2S+1 = 1$) and largest ($2S+1 = 19$) spin states. We summarize such gaps for the different geometries in Tab.~\ref{tab:SpinGapGeoms}, in which we compute the high spin energies with ASCI using the CASSCF optimized orbitals for the corresponding singlet state. We further report in parentheses the equivalent gaps computed in the localized ROHF basis with DMRG. While the high-spin state is generally higher in energy, we observe a strong geometry dependence for this gap, in several cases well resolved with the accuracy of our methods. In particular, the 2A geometry presents the significantly smallest gap, of only $\sim 1$~mHa, approximately one order of magnitude smaller than the gap for the other geometries. This is significant, since geometry 2A was optimized for the high spin state. While these trends can be observed in both the localized ROHF and optimized CASSCF orbital bases, the orbital optimization reduces the gaps by effectively a factor of two.

Comparing the CASSCF results in Tab.~\ref{tab:MCSCFFe4} with the DMRG energies in the ROHF basis from Tab.~\ref{tab:MB_ROHF_Fe4}, we note a significant energy stabilization. Indeed, in the 2R geometry we observe over 100 mHa energy difference between the extrapolated results before and after the CASSCF optimization. Including the bridging and ligand S-\textit{p} orbitals is capturing a relevant component of the correlation energy, supported by the double-exchange~\cite{Zener1951,Anderson1955,deGennes1960} picture which is used to motivate the spin structure in these clusters. Below we study the correlation energy as a function of the active space size, which will further strengthen this interpretation. The effective energy gap between the reference geometry 2R and geometries 2A-2C increases upon orbital optimization, while the gaps between spin states are similar at the DMRG level. 

Beyond just considering the energetics, it is interesting to investigate the change of character of the active space orbitals upon the CASSCF optimization. In Fig.~\ref{fig:2A_sing_orbrot} we show sample orbitals before and after the CASSCF optimization for the 2A geometry, and $2S+1 = 1$ spin state. We observe that there are still 20 orbitals of essentially exclusive Fe-3\textit{d} character, c.f. the two lower left panel in Fig.~\ref{fig:2A_sing_orbrot}. However, these present at times some degree of pairing into Fe-dimers. By this we mean that some of these orbitals are linear combinations of Fe-3\textit{d} localized on two Fe centers. For instance, note how in the lower left panel of Fig.~\ref{fig:2A_sing_orbrot}, besides a dominant Fe-\textit{d} contribution in one of the four iron centers, there is a minor yet significant Fe-\textit{d} contribution from another iron center, bridged by the connecting S-\textit{p} orbitals. Further, for geometries where this pairing is strongly present, it defines two clear pairs, i.e. we only observe pairing between Fe-3\textit{d} orbitals in Fe atom pairs 1-2 and 3-4, but never between 3 and 1 or 3 and 2. These pairs coincide with the shortest Fe--Fe distances in the corresponding geometry, c.f. Sec.~\ref{sec:GeomsAndDetails}.

\begin{figure}
    \centering
    \includegraphics[width=0.5\textwidth]{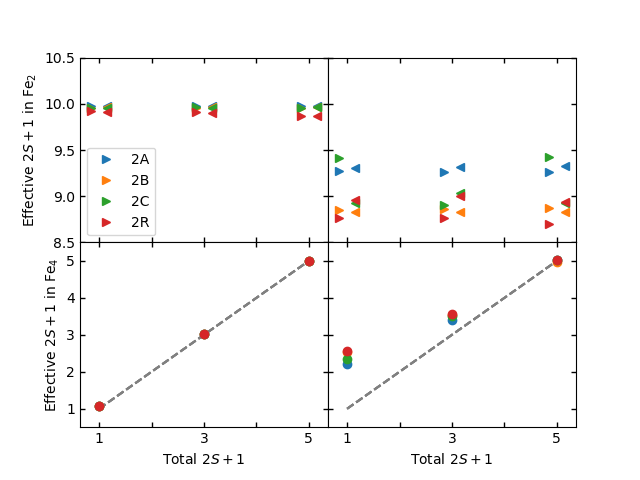}
    \caption{Effective spin multiplicity $2S+1$ for different subsets of Fe-\textit{d} orbitals from the CASSCF optimized SP-ASCI wave functions of the four cluster geometries with the (aug)-cc-pVDZ basis set and the (54e, 36o) active space. Upper panels: Multiplicities considering the Fe-\textit{d} orbitals of the two ferromagnetically coupled Fe-dimers within the [Fe$_4$S$_4$(SMe)$_4$]$^{-2}$ cluster. The left panel is computed in the CASSCF basis, the right panel is rotated to the localized ROHF basis. Lower panels: Multiplicities considering the Fe-\textit{d} orbitals of all four Fe atoms, in the CASSCF basis (left) and rotated to the localized ROHF basis (right). See text for details.}
    \label{fig:spin_corrs_geoms}
\end{figure}

This is a noteworthy phenomenon, since it is consistent with the magnetic structure expected for the low spin states of these systems: two high-spin Fe-dimers anti-ferromagnetically coupled to result in an overall singlet state. Still, single-particle orbitals are not physically well defined magnitudes. Thus, we confirm this picture by computing actual observables, such as spin-spin correlation functions ${C^{S}_{a,b} = \langle \vec{S}_a \cdot \vec{S}_b\rangle_0}$, accessible through the 2-RDM, where $a$ and $b$ denote single-particle orbitals and $\langle\cdot\rangle_0$ denotes a ground state expectation value. From these correlation functions, we can evaluate effective spin multiplicities for the \textit{d}-orbitals of the two Fe-dimers present in each [Fe$_4$S$_4$(SMe)$_4$]$^{-2}$ cluster. These effective multiplicities are shown in the upper panels of Fig.~\ref{fig:spin_corrs_geoms}, with two triangle markers of different orientation corresponding to each one of the two dimers in each geometry. The 2-RDM's were computed from the CASSCF optimized SP-ASCI wave functions with 5 million determinants. The left panels show effective multiplicities in the CASSCF orbital basis, the right panels in the localized ROHF orbital basis (i.e. the 2-RDM was rotated back to the localized ROHF basis). In the CASSCF orbital basis (left panels), we observe high and equal effective multiplicities of $2S_{eff}+1\sim10$ for all Fe-dimers in all geometries and all total spin states. This corresponds, in its simplest interpretation, to 9 unpaired electrons of parallel spin. Each dimer is composed of two ferromagnetically coupled Fe atoms. Since this is independent of the total spin state of the cluster, i.e. the Fe-dimers are high-spin for all singlet, triplet and quintet, this suggests a weak ligand field splitting for the d-orbitals of the Fe-centers, as well as a dominant double-exchange mechanism producing ferromagnetic order.  When considering then the effective multiplicity due to the \textit{d}-orbitals of all four Fe atoms (lower panels in Fig.~\ref{fig:spin_corrs_geoms}), we see that in the CASSCF orbital basis (left panel) the Fe-\textit{d} orbitals account for the full cluster spin, showing that the high-spin Fe-dimers couples antiferromagnetically with different relative orientations to give the total spin states. Rotating the 2-RDMs to the localized ROHF basis, and recomputing the effective dimer and [Fe$_4$S$_4$(SMe)$_4$]$^{-2}$ 
mutiplicities in terms of the localized Fe-\textit{d} orbitals, the picture changes slightly, see right panels in Fig.~\ref{fig:spin_corrs_geoms}. Here, the Fe-dimers have a slightly reduced effective multiplicity, though still possessing a high-spin indicative of ferromagnetic correlation, and the full Fe-\textit{d} orbital manifold does not account for the total spin state of the cubane cluster. These changes suggest the presence of spin fluctuations from the Fe-\textit{d} orbitals into the rest of the system, likely the S-\textit{p} orbitals. 

We can also observe similar fluctuations between the Fe-\textit{d} and S-\textit{p} orbitals in the orbital charge density, by examining the diagonal terms of the 1-RDM. In the lower panel of Fig.~\ref{fig:2A_sing_occs}, we show the orbital charge densities for the Fe-3\textit{s}/3\textit{p}, as well as valence S-\textit{p} and Fe-\textit{d} orbitals for the 2A cluster geometry, singlet state. These correspond to the diagonal 1-RDM components, rotated back to the localized ROHF basis. The orbital charge densities in the optimized CASSCF molecular orbitals are shown in the upper panel of the same figure. As the figure shows, there is clear charge density fluctuations from the localized Fe-\textit{d} orbitals into the S-\{textit{p}. 

\begin{figure}
    \centering
    \includegraphics[width=0.5\textwidth]{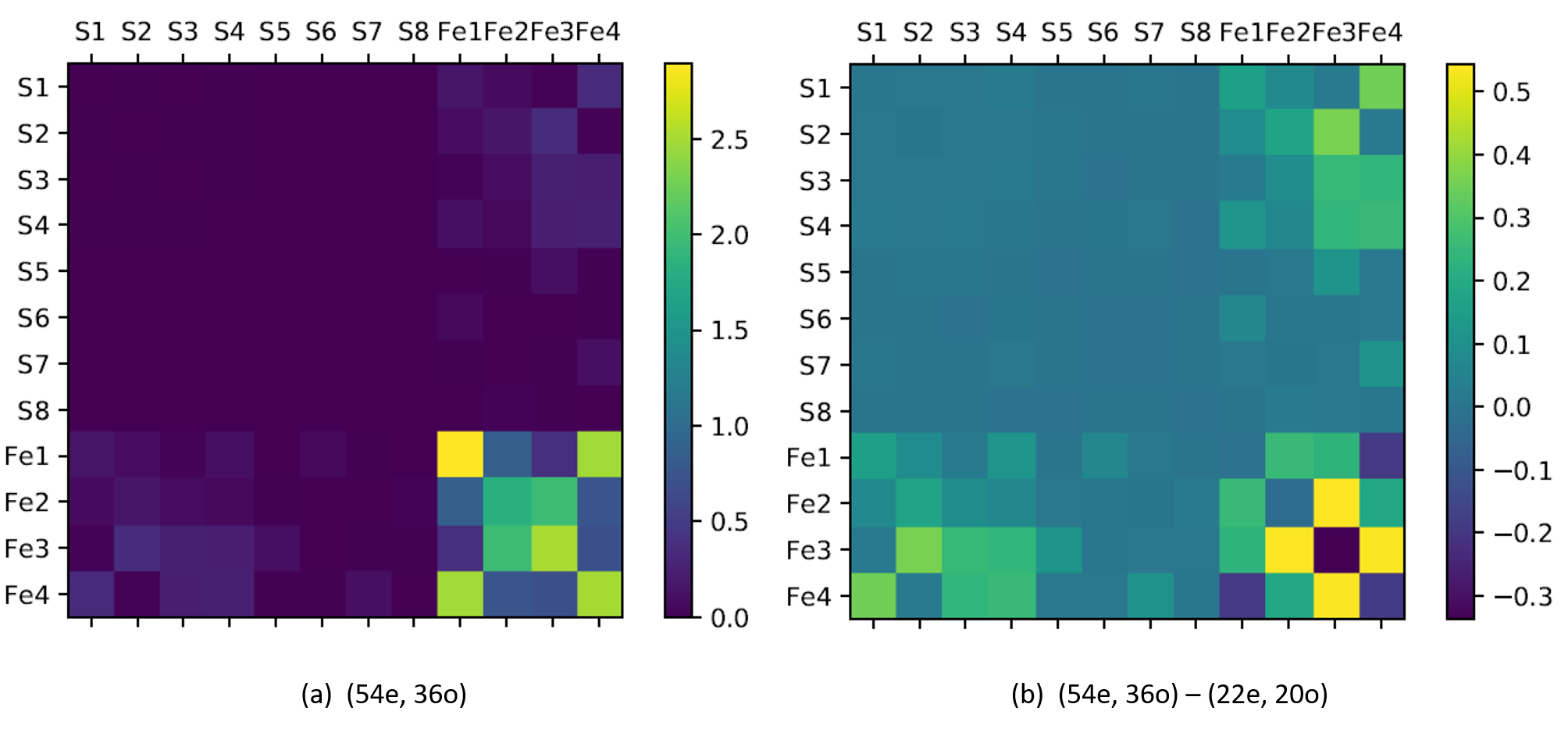}
    \caption{Mutual information for the 2A cluster, singlet state, between different orbital groups. Fe$_i$ corresponds to the valence \textit{d}-orbitals of the $i$-th Fe atom, while S$_i$ refers to the valence \textit{p}-orbitals of the $i$-th S atom. The left panel shows the mutual information computed from the ROHF localized orbitals using the (54e, 36o) active space, while the right panel shows the difference between the results using the (54e, 36o) and (22e, 20o) active spaces.}
    \label{fig:mutualInfo}
\end{figure}

To further support the double-exchange mechanism, we have additionally studied pair-wise orbital correlations by means of the mutual information. \cite{legeza_review} Since the mutual information is a two-orbital (rather than a two-electron) quantity, it includes certain elements of 4-point correlation functions (i.e. ground state expectation values including up to four pairs of creation/annihilation operators). \cite{boguslawski_2014} We considered Fe-\textit{d} orbitals grouped by Fe-atom, as well as S-\textit{p} orbitals grouped by S-atom, and show the mutual information between these groups in Fig.~\ref{fig:mutualInfo}. We show the mutual information between these groups from the 2A DMRG singlet wave function in the ROHF localized basis, using the (54e, 36o) active space in the left panel. We observe clear signatures of correlation between the Fe1-Fe4 and Fe2-Fe3 dimers, which correspond to the ferromagnetically pairs identified with the spin-spin correlation functions in Fig.~\ref{fig:spin_corrs_geoms}. We further observe some minor, though noticeable degree of correlation between the Fe-\textit{d} and bridge S-\textit{p} (S1-S4) orbitals, while the ligand S-\textit{p} (S5-S8) orbitals have weaker correlation to the Fe-\textit{d}'s. The right panel of Fig.~\ref{fig:mutualInfo} shows the difference between this mutual information computed in the (54e, 36o) active space, and the mutual information from a (22e, 20o) active space calculation. This active space does not include the S-orbitals, and thus there is no correlation between them and the Fe-\textit{d}'s. The difference figure thus reveals the effect of the S-\textit{p} orbitals in the Fe-Fe orbital correlations. We observe that including the S-\textit{p} orbitals explicitly changes the Fe-Fe correlations significantly. In particular, it is interesting to note that the correction to the dimer Fe1-Fe4 and Fe2-Fe3 correlations have opposite signs. Since the resulting mutual information (see right panel) are comparable for both panels, this may be the consequence of asymmetries in the localized ROHF basis. This analysis of mutual information, as well as the previous discussion in terms of spin-spin correlation functions and charge densities, is in perfect agreement with the magnetic structure expected for these clusters, and further with an underlying double-exchange interaction between Fe atoms, mediated by S electrons.

Beyond the Fe-3\textit{d} pairing, the majority of the remaining active space orbitals show mixed character between bridge S-\textit{p} orbitals and Fe-3\textit{d}, further supporting the double-exchange picture, c.f. lower right panel of Fig.~\ref{fig:2A_sing_orbrot}. Observing both these features, Fe-\textit{d}/Fe-\textit{d} pairing and Fe-\textit{d}/S-\textit{p} bridging, emerging from the CASSCF optimization is an important indicative that the optimized orbitals are a fundamentally better basis to describe the electronic properties of the iron-sulfur clusters.

However, not all optimized orbitals in the active space follow the previous scheme as nicely as shown in Fig.~\ref{fig:2A_sing_orbrot}. In particular, throughout all geometries and spin states, several of the S-\textit{p} and Fe-\textit{d} orbitals in the active space get substituted by Fe-\textit{p} orbitals during the CASSCF calculations. These orbitals are nonetheless otherwise unmixed with the rest of the active space constituents, and given their non-valence character seem unlikely to play a relevant role in the reactive properties of these clusters. To discern whether the inclusion of Fe-\textit{p} orbitals is an artifact of the optimization, or actually important from a physical point of view, we consider CASSCF calculations on the 2A geometry with various active space sizes in the next section.

\subsection{The Effect of the Active Space}

Whether CASSCF captures the correct physical behavior can be active space dependent, especially in strongly correlated systems such as iron-based clusters.~\cite{Levine2020} Therefore, we consider in this subsection the effect of the active space choice in the particularly challenging 
[Fe$_4$S$_4$(SMe)$_4$]$^{-2}$ clusters. To this end, the following active spaces of increasing size were considered:

\begin{itemize}
    \item (22e, 20o): This is a minimal active space, containing exclusively the Fe-\textit{d} orbitals and electrons. While the Fe electrons are likely the main actors in the catalytic properties of the cluster, this active space does not account explicitly for charge or spin fluctuations between the iron and sulfur centers, and thus can only account implicitly for the double-exchange mechanism which typically governs the magnetic correlations in transition metal clusters.~\cite{Sharma2020} Still, recent studies~\cite{Li2020} have shown that this type of minimal active space may be enough to capture energy gaps, and thus we include it in our study. It further offers an important point of reference to infer the role of the additional orbitals included in the subsequent active spaces.
    \item (46e, 32o): This active space includes the Fe-\textit{d} orbitals and the S-\textit{p} orbitals of the four bridging S atoms. Thus, it is the minimal active space to explicitly account for double-exchange interactions.
    \item (54e, 36o): This active space further includes the four S-\textit{p} orbitals from the ligand S atoms, pointing towards their bonded Fe center. This is the same active space considered in Ref.~\cite{Sharma2014}, and accounts for ligand effects into the iron-sulfur cluster. Given the relatively small size of the cluster, including only four Fe centers, these ligand effects are likely to be important for a quantitative description of the system, and they are known experimentally to change photo-emission spectra appreciably in the valence region of iron-sulfur cubanes.~\cite{Wang2003, Waters2007}
    \item (86e, 52o): In the previous section, we have observed that performing a CASSCF calculation in the (54e, 36o) active space resulted in some of the active orbitals being exchanged for the Fe-\textit{p} ones. Thus, in this active space we include the twelve Fe-3\textit{p} and four Fe-3\textit{s} orbitals. 
\end{itemize}

We concentrate here on the 2A geometry, which our previous results show as the most variationally stable, both at the mean-field and correlated treatments. Given the previous discussion on the anti-ferromagnetic coupling between high-spin Fe-dimers, it is likely that the 2A geometry is the most stable because it has the shortest relative Fe--Fe dimer bond length within the high spin dimers, as discussed in Sec.~\ref{sec:GeomsAndDetails}. In this cluster, the high-spin dimers present a bond length $\sim11\%$ smaller than the others Fe--Fe bonds, while for 2B-R the high spin dimers show only $\sim3\%$ shorter bonds. 

For all the active spaces described above, we perform CASSCF calculations with the SP-ASCI solver following the exact same procedure as with the (54e, 36o) active space in the previous section, starting from Pepek-Mezey localized HS-ROHF orbitals. We report the CASSCF and ASCI+PT2 extrapolated energies for the different active spaces, and three lowest lying spin states for cluster geometry 2A in Tab.~\ref{tab:MCSCF2A_ASs}.

\begin{table}[hbt!]
  \begin{center}
    \begin{tabular}{|l|c|c|c|c|c|} 
    \hline
    Act. Space & $2S + 1$ & CASSCF & ASCI+PT2 extrapol. \\
        &          & SP-ASCI-SCF & SP-ASCI \\
    \hline
    (22e, 20o) & 5 & -9.088122 & -9.088360 $\pm$ 0.000037 \\
      & 3 & -9.088318 & -9.088575 $\pm$ 0.000002 \\
      & 1 & -9.088439 & -9.088638 $\pm$ 0.000010 \\
    \hline
    (46e, 32o) & 5 & -9.142546 & -9.148286 $\pm$ 0.000524 \\
      & 3 & -9.142428 & -9.147207 $\pm$ 0.000692 \\
      & 1 & -9.142759 & -9.147001 $\pm$ 0.000425 \\
    \hline
    (54e, 36o) & 5 & -9.150451 & -9.155331 $\pm$ 0.000981 \\
      & 3 & -9.150856 & -9.155357 $\pm$ 0.000843 \\
      & 1 & -9.150950 & -9.155423 $\pm$ 0.000681 \\
    \hline
    (86e, 52o) & 5 & -9.171896 & -9.178860 $\pm$ 0.000497 \\
      & 3 & -9.171579 & -9.178939 $\pm$ 0.000727 \\
      & 1 & -9.171783 & -9.178807 $\pm$ 0.000730 \\
    \hline
    \end{tabular}
  \end{center}
  \caption{CASSCF and extrapolated energies ($E - 8380.0$ Ha) using SP-ASCI different active spaces for the 2A geometry of the
  [Fe$_4$S$_4$(SMe)$_4$]$^{-2}$ cluster, starting from the high spin ($2S + 1 = 19$) ROHF orbitals, localized with the Pipek-Mezey scheme. All calculations are performed with the mixed aug-cc-pVDZ (Fe,S) cc-pCDZ (C,H) basis set. The CASSCF energies correspond to SP-ASCI calculations with $5\cdot10^6$ determinants. These correspond to the final step of a series of SP-ASCI-SCF calculations starting at $1\cdot10^5$ determinants, and progressively increasing the number of determinants to improve the orbitals sequentially. The ASCI+PT2 extrapolated results are estimating the FCI limit, using a linear extrapolation from calculations with $5\cdot10^5$, $1\cdot10^6$, $2\cdot10^6$, $5\cdot10^6$ determinants, starting from the orbitals obtained from the SP-ASCI-SCF with $5\cdot10^6$ determinants. The (22e, 20o) extrapolation converged by the $2\cdot10^6$ determinant calculation, the (86e, 52o) one was extended to $7\cdot10^6$ determinants. The error bars correspond to the standard deviation of the linear fit, and are thus just a measure of the extrapolation error alone. For the quintet (86e,~52o) extrapolation, the $10^6$ determinant calculation was disregarded as an outlier.}
  \label{tab:MCSCF2A_ASs}
\end{table}

\begin{figure}
    \centering
    \includegraphics[width=0.5\textwidth]{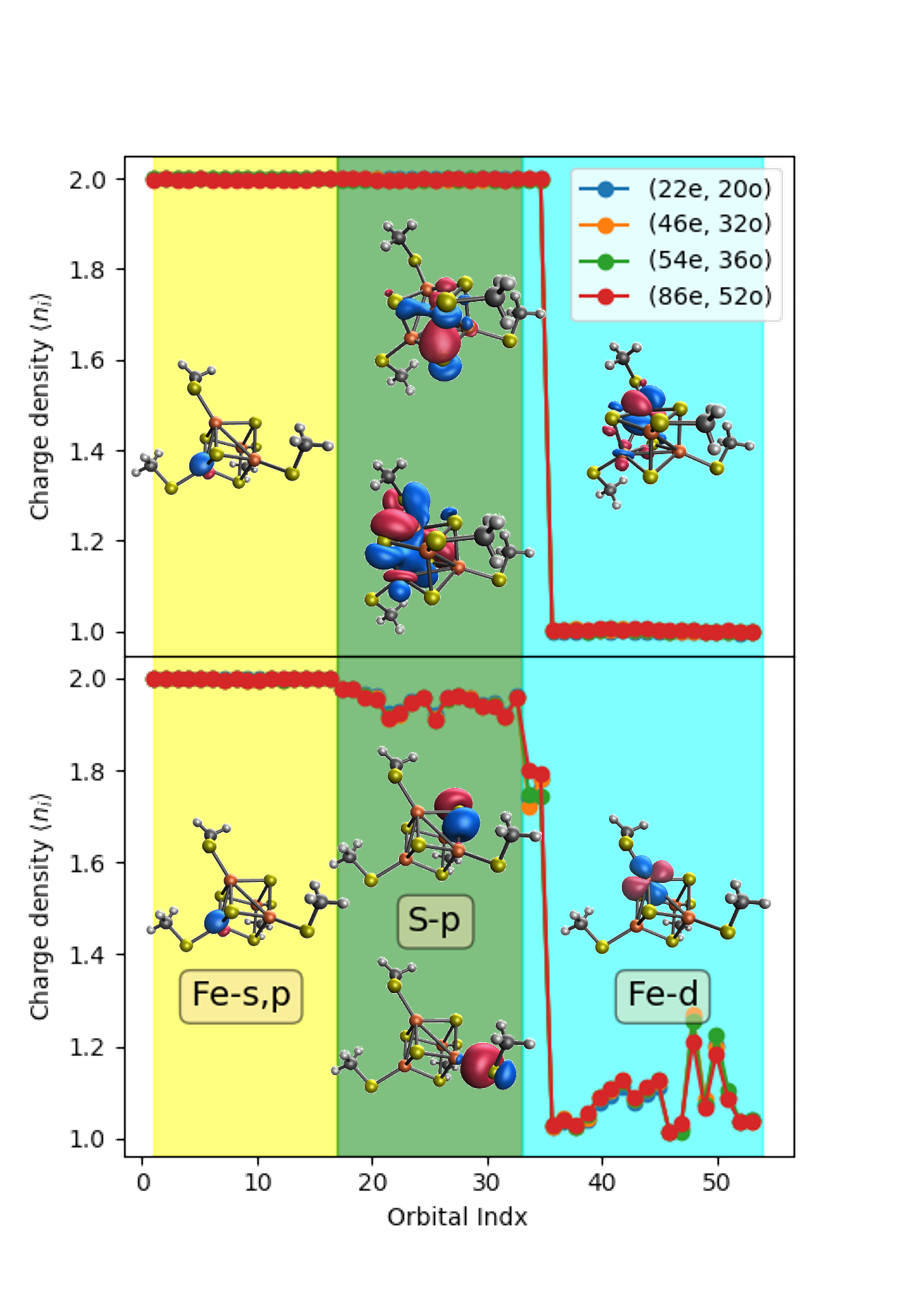}
    \caption{Upper panel: Orbital charge densities in the CASSCF wave function for the 2A geometry of the 
    [Fe$_4$S$_4$(SMe)$_4$]$^{-2}$ cluster, $2S+1 = 1$, for different active spaces in the optimized orbital basis. The wave function is computed with SP-ASCI, and $5\cdot10^6$ determinants. We label the orbitals by their character before the optimization, i.e. the labels correspond to the original active space order fed into the CASSCF routine. Lower panel: Same orbital charge densities of same wave function, in the localized ROHF basis. See text for details.}
    \label{fig:2A_sing_occs}
\end{figure}

Examining the CASSCF energies in Tab.~\ref{tab:MCSCF2A_ASs}, it becomes clear that the minimal active space (22e, 20o) misses, as expected, a significant amount of the correlation energy, having a gap with respect to the next active space (46e,~32o) of $\sim50$~mHa. Furthermore, compared with the high-spin ROHF energy in Tab.~\ref{tab:key}, the (22e,~20o) active space is only $\sim5$ mHa lower in energy (we observe a similar behavior in the Fe$_3$ results with the (15e,~15o) active space in the ESI). Upon further increase of the active space, we observe additional stabilization energies: the ligand S-\textit{p} orbitals recover $\sim10$~mHa, and the Fe-\textit{p} and Fe-s orbitals surprisingly accounting for an additional $\sim20$~mHa. The fact that the Fe-\textit{s},\textit{p} orbitals account for a comparable amount of correlation energy than the ligand S-\textit{p} indicates that an accurate treatment of the electronic structure requires both sets, suggesting that larger active spaces than are usually considered are likely key to accurately predicting the electrochemical properties of iron-sulfur systems. Similar conclusions have been drawn from single-point selective CI and DMRG calculations on the FeMoco cofactor~\cite{Li2019}. This notion is further supported by the observation, noted above, that optimized orbitals in both the (46e, 32o) and (54e, 36o) active spaces end up including some Fe-\textit{p} orbitals after the ASCI-SCF calculation.

\begin{figure}
\centering
\includegraphics[width=0.5\textwidth]{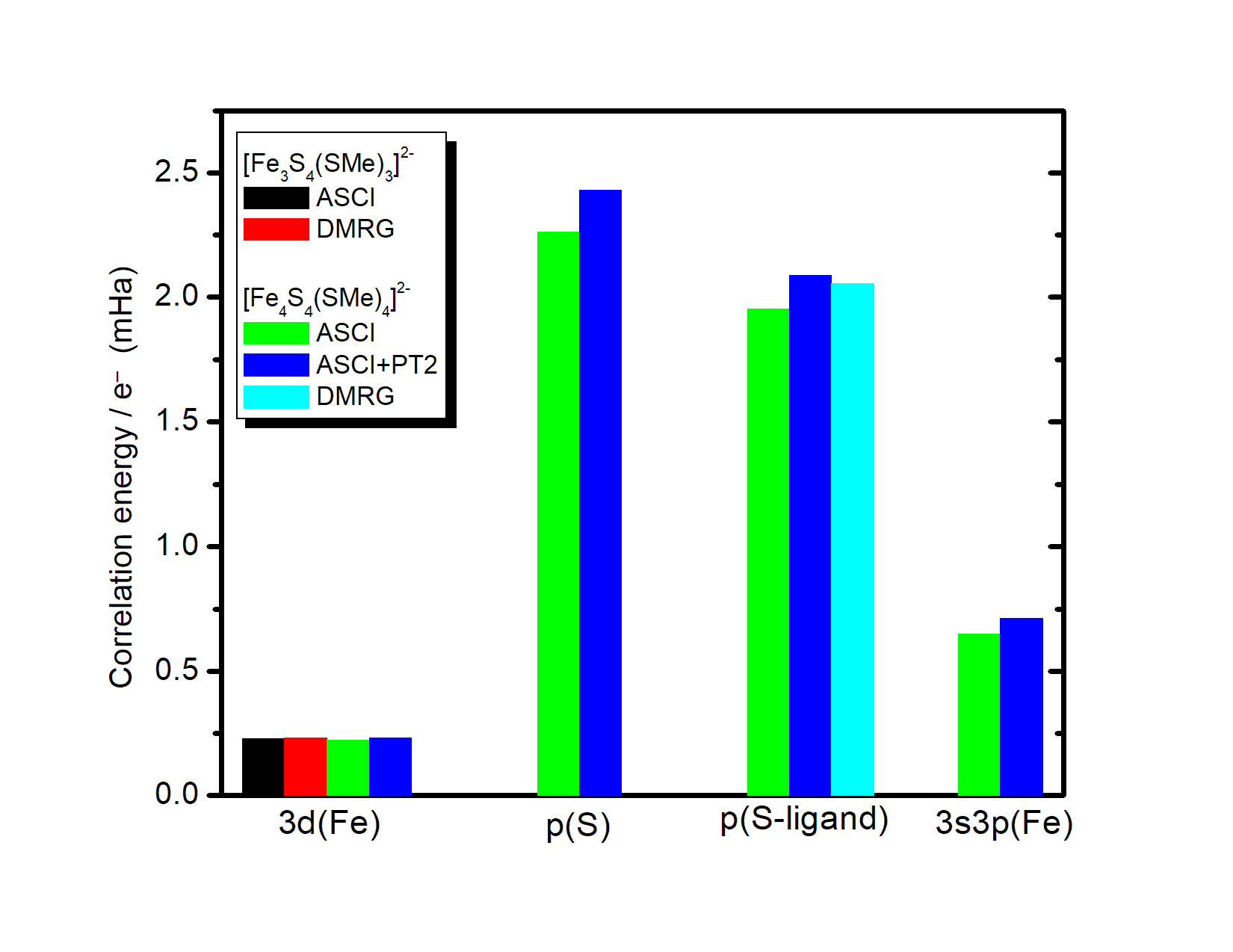}
\caption{Correlation energy of Fe-\textit{d}, S-\textit{p}, and Fe-3\textit{s}3\textit{p} electrons in the 2A geometry of the [Fe$_4$S$_4$(SMe)$_4$]$^{-2}$ cluster, and of Fe-\textit{d} orbitals for the 1A geometry of the [Fe$_3$S$_4$(SMe)$_3$]$^{-2}$ cluster.}
\label{fig:corr_ens} 
\end{figure}

In Fig~\ref{fig:corr_ens} we summarize the correlation energy per electron obtained by including each new set of orbitals into the correlated CASSCF SP-ASCI calculation for the 2A cluster geometry, as well as the correlation energy per electron in the Fe-\textit{d} orbitals for one of the [Fe$_3$S$_4$(SMe)$_3$]$^{-2}$ clusters included in the ESI~\cite{supp}. As mentioned before, the correlation energy resulting from the \textit{d} electrons is very small, about~0.2 mHa /e- at ACSI and DMRG, see Fig. ~\ref{fig:corr_ens}, for both clusters. This is the reason why the states are energetically degenerate for different spin multiplicities. 
The correlation energy results mainly from the \textit{p} electrons of S and it is significantly larger than the \textit{d} electrons of Fe, i.e., ~2.5~mHa~/~e- for the \textit{p} electron of S bridge and ~2.0~mHa~/~e- for \textit{p} electron of S ligand. It is interesting that the correlation of the 3\textit{s} and 3\textit{p} electrons of Fe are 0.7~mHa~/~e-. The correlation that stems from S-\textit{p} orbitals can be attributed to two complementary reasons: i) the S-\textit{p} orbitals are essentially doubly occupied, but there is an interaction with the low-lying empty 4\textit{s} orbitals of the Fe$^{+2}$($^5$D, 3d$^6$) and Fe$^{3+}$($^6$S, 3d$^5$) (see SI~\cite{supp}); and ii) the double-exchange interaction of half occupied Fe-\textit{d} and S-\textit{p} orbitals, where both the Fe-\textit{d} and S-\textit{p} orbitals need to be included in the active space in order to account for the correlation energy. Finally, the correlation due to the Fe-(3\textit{p}/3\textit{s}) orbitals results from the fact that these orbitals have the right symmetry to couple to both the Fe-\textit{d} orbitals and with the empty Fe-4\textit{s} ones.

The previous energetic considerations thus seem to suggest that it is necessary to include all S-\textit{p}, and even further the Fe-\textit{p}/\textit{s} orbitals into the active space for an accurate description of ground state of these iron-sulfur clusters. Still, when examining the charge densities of these CASSCF optimized orbitals for the different active spaces (i.e. the diagonal elements of the 1-RDMs), represented in the upper panel of Fig.~\ref{fig:2A_sing_occs} for the 2A geometry in the singlet state, we see that even when treated explicitly, the Fe-\textit{s},\textit{p} orbitals, and indeed the S-\textit{p}/Fe-\textit{d} molecular orbitals, are effectively inactive, i.e. consistently doubly occupied in the wave function. Rotating the corresponding 1-RDM into the localized ROHF basis, we can compute the charge densities of the atomic Fe-\textit{s}/\textit{p}/\textit{d} and S-\textit{p} orbitals, shown in the lower panel of Fig.~\ref{fig:2A_sing_occs}. While this unveils some degree of charge transfer between the Fe-\textit{d} and S-\textit{p} orbitals further supporting a double-exchange mechanism, the Fe-\textit{s}/\textit{p} orbitals remain essentially inactive, with no appreciable charge fluctuations away from them. For the Fe-\textit{s}/\textit{p} orbitals this suggests that the origin of the energy lowering upon their inclusion in the active space may arise due to an improvement in the orbital optimization process.

\begin{figure}
    \centering
    \includegraphics[width=0.5\textwidth]{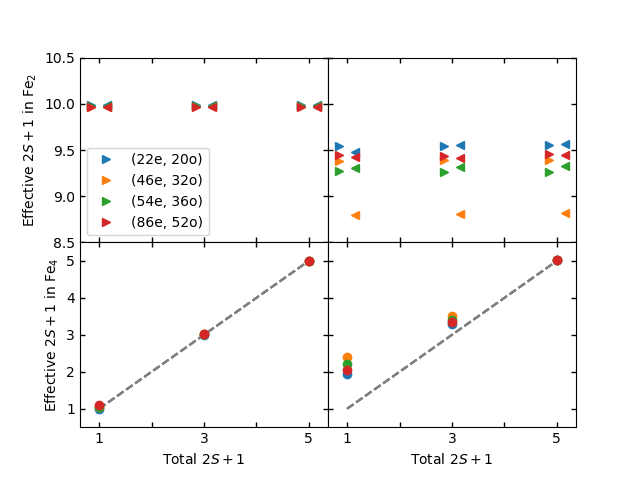}
    \caption{Effective spin multiplicity $2S+1$ for different subsets of Fe-\textit{d} orbitals from the CASSCF optimized SP-ASCI wave functions of the 2A geometry in the (aug)-cc-pVDZ basis set and different active spaces. Upper panels: Multiplicities considering the Fe-\textit{d} orbitals of the two ferromagnetically coupled Fe-dimers within the Fe4 cluster. The left panel is computed in the CASSCF basis, the right panel is rotated to the localized ROHF basis. Lower panels: Multiplicities considering the Fe-\textit{d} orbitals of all four Fe atoms, in the CASSCF basis (left) and rotated to the localized ROHF basis (right). See text for details.}
    \label{fig:spin_corrs_actspac}
\end{figure}

Like in the previous section, we examine the spin-spin correlation functions within the Fe-\textit{d} orbitals for the 2A geometry singlet, triplet and quintet states as a function of the active space size in Fig.~\ref{fig:spin_corrs_actspac}. In the basis of CASSCF optimized orbitals (left panels), we see that the picture discussed in Fig.~\ref{fig:spin_corrs_geoms} is independent of the size of the active space. Even the small (22e,~20o) active space, which only includes Fe-\textit{d} orbitals explicitly, can capture the organization of the four Fe atoms into two high-spin Fe-dimers which couple anti-ferromagnetically to account for, essentially, the full spin state of the cluster. When rotating the correlation functions back into the localized ROHF basis, we once again observe some degree of spin fluctuation away from the Fe-\textit{d} orbitals: the Fe-dimers have lower effective spin multiplicity (upper right panel in Fig.~\ref{fig:spin_corrs_actspac}), and the Fe-\textit{d} orbitals do not account for the spin state of the full cluster (lower right panel in Fig.~\ref{fig:spin_corrs_actspac}). The results are essentially independent on the active space, except for the fact that in the (46e,~32o) active space, the two high-spin Fe-dimers show different effective multiplicities.

From the present discussion, we see that the smallest active space (22e,~20o) describes static properties, such as charge densities and spin-spin correlations, equally well than more complex active spaces including S-\textit{p} and Fe-\textit{s}/\textit{p} orbitals. The orbital optimization in the CASSCF procedure seems to compensate for the lack in active orbitals. Still, considering the change in correlation energy by active space, see Fig.~\ref{fig:corr_ens}, the Fe-\textit{d} orbitals alone cannot account for anything but a minute part of the correlation, and it is necessary to include all Fe-\textit{d}, S-\textit{p}, and apparently Fe-\textit{s}/\textit{p}, in the active space to capture the correlation energy. These observations seem to contradict each other. A possible reconciliation would be to check whether the correlation energy due to the S-\textit{p} and Fe-\textit{s}/\textit{p} orbitals can be recovered perturbatively from the (22e,~20o) ground state wave function. This would explain both the consistent static properties across active spaces, as well as the sizeable correlation energy due to the S-\textit{p} and Fe-\textit{s}/\textit{p} orbitals. We checked this by computing the ASCI+PT2 energies for the 2A singlet state starting from the (22e,~20o) active space wave function with 2 million determinants, but including the S-\textit{p} and Fe-\textit{s}/\textit{p} orbitals for the perturbative correction. However, this could only account for $\sim$5 mHa, which is only $\sim10\%$ of the total correlation energies in Fig.~\ref{fig:corr_ens}. The S-\textit{p} and Fe-\textit{s}/\textit{p} orbitals seem to be needed explicitly in the active space in order to capture the correlation energy accurately. Our results suggest that static properties, within and without the active space, can be captured in the CASSCF orbital rotation, while dynamical information such as correlation energies require an explicit description within the active space.

\begin{table}[hbt!]
  \begin{center}
    \begin{tabular}{|l|c|c|c|c|c|} 
    \hline
    Act. Space & $2S + 1 = 1$ & $2S + 1 = 19$ & Spin Gap [mHa] \\
    \hline
    (22e, 20o) & -9.088439 & -9.082663 & -5.776 \\
    (46e, 32o) & -9.142759 & -9.140243 & -2.516 \\
    (54e, 36o) & -9.150950 & -9.149944 & -1.006 \\
    (86e, 52o) & -9.171783 & -9.173704 & 1.921 \\
    \hline
    \end{tabular}
  \end{center}
  \caption{ASCI energies ($E - 8380.0$ Ha) using SP-ASCI as solver for the 2A cluster geometry in the $2S+1 = 1$ and $2S+1 = 19$ spin states for different active space sizes. The energies are computed using the CASSCF orbitals optimized for the singlet state. The last column reports the spin gap in mHa, computed as $E_0^{2S+1 = 1} - E_0^{2S+1=19}$.}
  \label{tab:SpinGapAS}
\end{table}

Finally, it is interesting to note the progression of the energy gap between the smallest and largest spin states ($2S+1 = 1$ and $2S+1=19$ respectively) as a function of the active space, which we summarize in Tab.~\ref{tab:SpinGapAS}. The energies of the high spin state are obtained using the CASSCF optimized orbitals for the singlet state, with enough determinants to converge the energy to sub mHa accuracy ($5\cdot10^5$ for all active spaces instead of the largest one, which needed $10^6$). Optimizing the orbitals for the high spin state explicitly does not change the energies significantly (sub mHa differences). We observe a monotonic decrease in the spin gap with increasing active space, up to a change in sign in the (86e, 52o) resulting in the high spin state being lower in energy than all low-spin states. While the gaps are still too small to make any definite claim, it is encouraging that we observe the high spin to be the most stable state in the large active space simulation, especially since this is the spin state for which the 2A geometry was optimized (see above). This is a strong indication of the geometry dependence not only of the spin state gaps, but also of the spin state orderings in the Fe-S clusters.

\section{Conclusions}
\label{sec:conclusions}

In this work, we have for the first time used CASSCF with large active spaces to study model systems for Fe-based catalytic centers, namely the [Fe$_4$S$_4$(SMe)$_4$]$^{-2}$ cubanes, paying special attention to the role of geometry and spin state in the system's properties. We have employed highly accurate and complementary correlated solvers for the CASSCF problem, namely ASCI and DMRG. The excellent agreement between these two techniques gives us confidence in our results, and further shows that it is now possible to treat these highly complex systems with sophisticated many-body approaches such as CASSCF. Moreover, we have introduced the SP-ASCI approach, which proposes Hilbert space truncations preserving spin conservation by modifying the ASCI search in terms of CSF families. While SP-ASCI shows severe convergence issues in the mean-field single-particle basis studied, i.e. localized ROHF, using it as solver for CASSCF remedies this limitation and provides accurate results in excellent agreement with DMRG.

Our results show that orbital optimization improves the energies of different spin states significantly, reduces the spin energy gaps by a factor of two, and also brings to the forefront the underlying physical mechanism that dominates the system. Indeed, the optimized single-particle orbital basis is reminiscent of the double-exchange mechanism widely accepted to be responsible for the magnetic structure of these Fe-S clusters. Since single-particle orbitals are not physically well defined, we complement this interpretation with one-electron, two-electron and two-orbital correlation functions, namely charge densities, spin-spin correlations and mutual information. All these diagnostics support the double-exchange interpretation, and are consistent across different active spaces, geometries and spin states.

All geometries show essentially degenerate spin states, with gaps between successive spin states being of the order of 1~mHa or lower. This is explained by examining the correlation energy as a function of active space size, which shows that the Fe-\textit{d} electrons alone do not contribute to it significantly. Since the spin state is essentially determined by the Fe-\textit{d} electrons, their near-degeneracy is not unexpected. The correlation energy comes in comparable amounts from bridge S-\textit{p}, ligand S-\textit{p} as well as Fe-\textit{s}/\textit{p} orbitals. The non-trivial correlation that stems from these orbitals can be attributed to two complementary reasons: i) the S-\textit{p} orbitals are essentially doubly occupied, but there is an interaction with the low-lying energy empty 4\textit{s} orbitals of Fe(II) and Fe(III) and ii) the double-exchange interaction of half occupied Fe-\textit{d} and S-\textit{p} orbitals, where both the Fe-\textit{d} and S-\textit{p} orbitals are needed to be treated in the active space in order to account for the correlation energy. Finally, the correlation due to the Fe-(3\textit{p}/3\textit{s}) orbitals results from the fact that these orbitals have the right symmetry to couple to the Fe-\textit{d} orbitals and with the empty Fe-4\textit{s} ones. To capture all this contributions accurately, it is necessary to include all these orbitals into the active space, as simple perturbative corrections on top of smaller active spaces do not seem capable of accounting for them.

While the small gaps between consecutive spin states makes a definite statement about detailed spin hierarchies difficult, we observe a significant geometry dependence of the largest spin gap, defined as the energy difference between the singlet and largest spin state. By arranging the Fe-atoms slightly asymmetrically, forming two dimers of $\sim$11$\%$ shorter bond length, this maximal gap is reduced by approximately one order of magnitude, and moreover may even invert its sign, stabilizing the high-spin state over the low spin manifold. It is this fine-tuning of spin-dependent energetics by subtle geometry changes that makes Fe-S based enzymes remarkable catalysts in biological systems. The significant reduction of spin energy gaps upon CASSCF orbital optimization is a strong indication that this type of sophisticated electronic structure treatment will prove crucial for an accurate description of the reactivity in these correlated systems. Finally, it has been reported recently that DFT-based methods are unreliable to predict the relative energy ordering of possible isomers of cofactors, such as the FeMoCo in nitrogenase~\cite{Cao2019Extremely}. Given their low computational cost, they are likely to remain as widely used methods to treat these complex systems. To this end, a useful and important scope of our work is to provide an accurate description of the challenging electronic structure of iron-sulfur cubanes that can be used as benchmark for DFT-based calculations.

\section*{Acknowledgements}
We wish to thank Dr. Simone Raugei of the Pacific Northwest National Laboratory for useful suggestions and discussions. DT, JB, LV, CMZ, SSX and WAJ acknowledge support from the Center for Scalable Predictive methods for Excitations and Correlated phenomena (SPEC), which is funded by the U.S. Department of Energy, Office of Science, Basic Energy Sciences, Chemical Sciences, Geosciences and Biosciences Division, as part of the Computational Chemical Sciences Program at Pacific Northwest National Laboratory. Battelle operates the Pacific Northwest National Laboratory for the U.S. Department of Energy. This manuscript has been authored by an author at Lawrence Berkeley National Laboratory under Contract No. DE-AC02-05CH11231 with the U.S. Department of Energy. 
NMT is grateful for support from NASA Ames Research Center and support from the AFRL Information Directorate under Grant  No.~F4HBKC4162G00. 
LV acknowledges support from Czech Science Foundation (grant no. 18-18940Y).
The calculations were performed as part of the XSEDE computational Project No. TG-MCA93S030. This research also used resources of the National Energy Research Scientific Computing Center, which is supported by the Office of Science of the U.S. Department of Energy under Contract No. DE-AC02-05CH11231 and resources of the Czech supercomputing center supported by the Czech Ministry of Education, Youth and Sports from the Large Infrastructures for Research, Experimental Development and Innovations project “IT4Innovations LM2015070.”

\section*{Author Contributions}
DT, WAJ and SSX proposed the original research; NT, LV, SSX and WAJ supervised different aspects of the project. DT performed the BS-DFT geometry optimizations and initial CASSCF calculations; CMZ performed the ASCI-SCF calculations and developed the spin-pure (SP) variant of ASCI; MM, JB and LV performed the DMRG calculations. DWY and NT contributed to the ASCI code. CMZ, DT and LV wrote the manuscript. All authors contributed to the analysis, discussion of the results, reviewed and modified the manuscript.

\section*{Conflicts of Interest}
The authors declare no conflicts of interest.
\bibliography{CMZ_refs, dmrg}

\end{document}